\documentclass[%
 reprint,
 amsmath,amssymb,
 aps,
]{revtex4-2}

\usepackage{graphicx}
\usepackage{dcolumn}
\usepackage{bm}
\usepackage{float} 
\usepackage{amsmath}
\usepackage{upgreek}
\usepackage{soul,xcolor}

\usepackage[normalem]{ulem}
\usepackage[per-mode=fraction]{siunitx}
\begin{document}
\setstcolor{red}

\title{Observation of the Wannier-Stark ladder in plasmonic waveguide arrays}

\author{Helene Wetter}
\author{Zlata Fedorova (Cherpakova)}%
\author{Stefan Linden}
 \email{linden@physik.uni-bonn.de}
\affiliation{Physikalisches Institut, Rheinische Friedrich-Wilhelms-Universit\"at Bonn, Nußallee 12, 53115 Bonn, Germany
}%

\date{\today}

\begin{abstract}
	Evanescently coupled waveguides are a powerful platform to study and visualize the wave dynamics in tight-binding systems. 
	Here, we investigate the propagation of surface plasmon polaritons  in arrays of dielectric loaded surface plasmon polariton waveguides with a propagation constant gradient acting as an effective external potential. Using leakage radiation microscopy, we 
	observe in real-space for single site excitation a periodic breathing of the wavepacket and an oscillatory motion in the case of Gaussian excitation of multiple waveguides.
	The corresponding momentum resolved spectra are composed of sets of equally spaced modes.
	We interpret these observation as the plasmonic analogues of Bloch oscillations and the Wannier-Stark ladder, respectively.
\end{abstract}

\maketitle
The solutions of the single-electron Schrödinger equation in a periodic potential take the form of extended Bloch states~\cite{bloch1929quantenmechanik}. Subjected to an additional spatially constant DC electric field and in the absence of scattering, the electronic wavefunctions localize. More specifically, the electrons perform an oscillatory motion known as Bloch oscillation, where the oscillation frequency $\omega_B$ is proportional to the applied field strength and the lattice period~\cite{zener1934theory}. The related spectral signature is the so-called Wannier-Stark ladder, i.e., the continuous band of Bloch states in the field-free case develops into a set of equally spaced energy levels with energy difference $\Delta E=\hbar \omega_B$ \cite{Wannier1960SolidStateTheory,mendez1993wannier}.

In bulk solids, Bloch oscillations and the formation of the Wannier-Stark ladder are inhibited by dephasing processes that typically happen on a time-scale significantly shorter than the achievable Bloch oscillation period duration $T_B=2 \pi / \omega_B$. This obstacle can be  overcome by employing an artificial super-lattice, for which the Bloch oscillation frequency $\omega_B$ is significantly increased due to the larger spatial period of the super-lattice. 
Following the pioneering experiments with semiconductor multiple quantum-well structures\,\cite{olbright1991optical,feldmann1992optical,voisin1988observation}, Bloch oscillations have been also observed in a number of other non-electronic systems, e.g., cold atoms in lattices\,\cite{wilkinson1996observation}, periodic dielectric films\,\cite{sapienza2003optical}, coupled optical waveguides\,\cite{pertsch1999optical},
THz acoustic phonons in coupled nanocavity structures\,\cite{lanzillotti2010bloch},
plasmonic systems\,\cite{block2014bloch}, exciton polaritons confined to coupled microcavity waveguides\,\cite{beierlein2021bloch}, and superconducting quantum processors\,\cite{guo2021observation}. 

\begin{figure}[t]
	\centering
	\includegraphics[width=\linewidth]{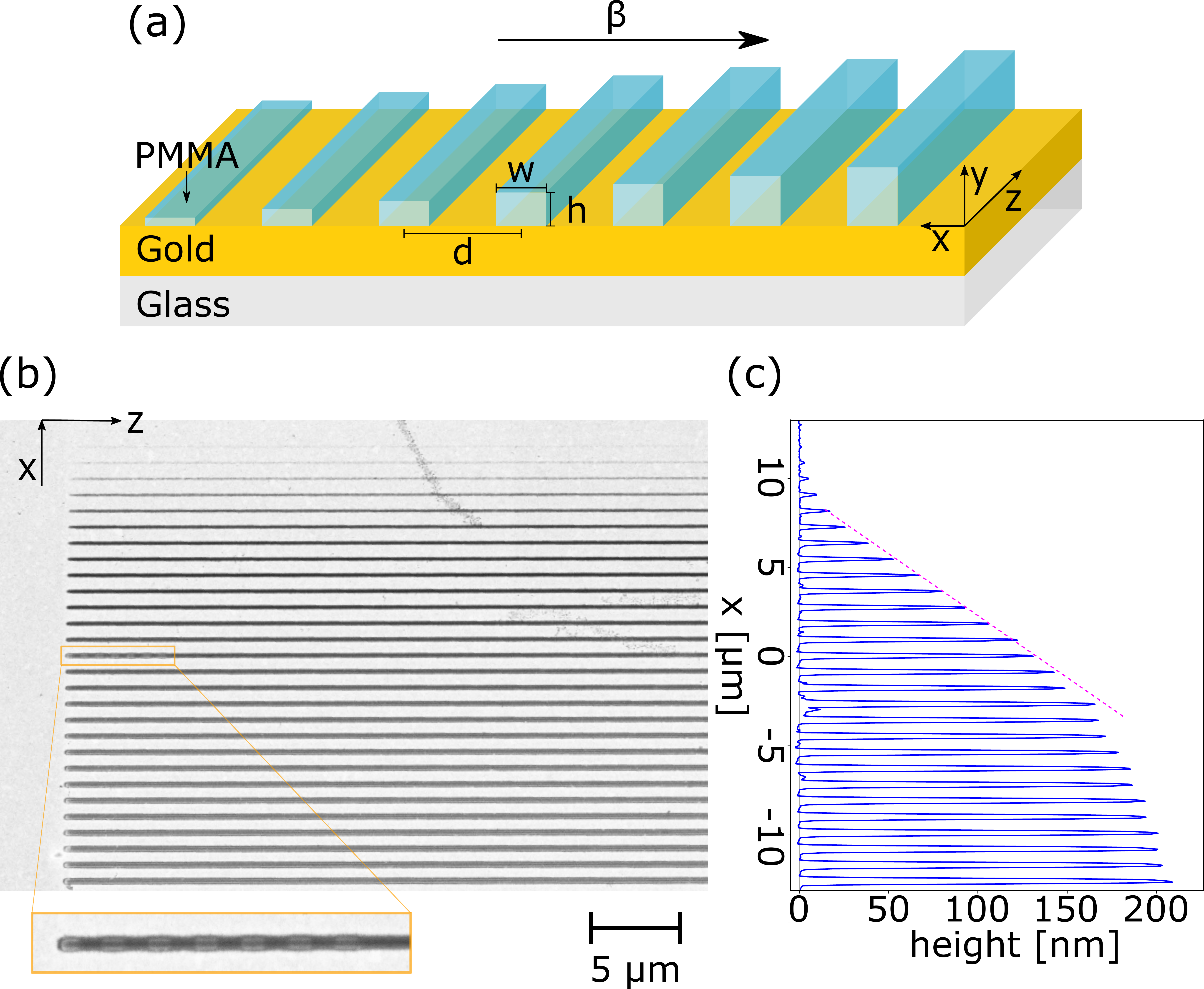}
	\caption{(a) Sketch of a DLSSPW array that mimics a lattice with applied electric field. The height gradient leads to a gradient of the propagation constant $\beta$.
	(b) Scanning electron micrograph of a DLSPPW array. A grating coupler deposited at the input of one of the waveguides is marked by the orange box. (c) Height profile of the DLSSPW array as recorded by atomic force microscopy.}
	\label{fig:afm_sem}
\end{figure}

In this letter, we report on the observation of the Wannier-Stark ladder and Bloch oscillations in arrays of evanescently coupled dielectric loaded surface plasmon polariton waveguides (DLSPPWs).
A gradient of the DLSSPW height mimics the  applied electric field (see Fig.\,\ref{fig:afm_sem} (a)).
Real- and Fourier space leakage radiation microscopy\,\cite{DREZET2008220} allows us to study the localization of the excited surface plasmon polariton (SPP)  wavepacket as well as the splitting of the continuous band into discrete states under the influence of the field.

The connection between the electron dynamics in a one-dimensional lattice and the spatial evolution of SPPs in an array of evanescently coupled waveguides is established by the so-called quantum-optical analogy~\cite{christodoulides2003discretizing,longhi2009quantum}. 
According to the coupled mode theory (CMT), the propagation of SPPs along an array of evanescently coupled DLSPPWs is governed by the following coupled set of equations:
\begin{align}
		\frac{\mathrm{d}a_m(z)}{\mathrm{d}z}=iC_{m-1,m}a_{m-1}(z) + i \beta_ma_m(z) + iC_{m,m+1}a_{m+1}(z). \label{CMT:equation}
\end{align}
Here, $a_m(z)$ is the amplitude of the SPP wave in the $m$-th waveguide at the position $z$, $C_{m,m-1}$ and $C_{m,m+1}$ are the coupling constants between the $m$-th  and the $m-1$-th and $m+1$-th waveguide,  respectively, and $\beta_m$ is the propagation constant of the $m$-th waveguide.
Interestingly, this equation of motion has the same mathematical form as the time-dependent Schr\"odinger equation in the tight-binding approximation.
A one-to-one comparison shows that the propagation distance $z$ in the optical system plays the role of time $t$, while the propagation constants $\beta_m$ and the coupling constants $C_{m,m+1}$ are related to the on-site energies and the hopping amplitudes, respectively, of the electronic system\,\cite{longhi2009quantum}.
Hence, we can map the time evolution of the probability density $\vert\Psi(x,t)\vert^2$ of a crystal electron in a one-dimensional lattice onto the spatial SPP intensity $I(m,z) \propto \vert a_m(z)\vert^2$ in the related waveguide array.
In order to simulate an electron in a periodic potential subjected to an additional DC electric field, the coupling constants and the propagation constants of the waveguide array should be chosen as $C_{m,m+1}=C$ and $\beta_m=\beta_0+ m \Delta \beta$, respectively, where $C$, $\beta_0$ and $\Delta \beta$ are constants for the given structure.
To meet the first condition ($C_{m,m+1}=C$), we choose for a given array the same center-to-center separation between all neighboring waveguides. 
A gradient of the propagation constant can be implemented, e.g., by varying the heights of the waveguides\,\cite{block2014bloch}. 

Figure\,\ref{fig:afm_sem} (a) depicts a scheme of the sample geometry. The DLSPPWs consist of poly(methyl methacrylate) (PMMA) ridges deposited on top of a glass substrate coated with 5 nm chromium as an adhesion layer and a 60 nm thick gold film. The width $w$ and height $h$ of the individual DLSPPWs as well as their arrangement in the array are defined by negative-tone gray-scale electron beam lithography (EBL)\,\cite{block2014bloch,cherpakova2017transverse}. In this process, we take advantage of the fact that the height of each DLSPPW can be controlled by the applied local electron dose.
A scanning electron micrograph of a typical sample is shown in Fig.\,\ref{fig:afm_sem} (b). 
This array consists of 29 DLSPPWs with a nominal width $w=300\,\mathrm{nm}$ that are arranged with a constant center-to-center distance $d=900\,\mathrm{nm}$ between the adjacent waveguides. 
The electron dose applied to the individual DLSSPWs was linearly increased from top 
(\SI{1,12}{\milli\coulomb\per\square\centi\meter}) to bottom (\SI{3,36}{\milli\coulomb\per\square\centi\meter}). 
The resulting height profile as determined by atomic force microscopy is presented in Fig.\,\ref{fig:afm_sem} (c). It confirms that in the central part of the array the waveguide heights and hence the propagation constants vary almost linearly along the $x$-direction.
In addition to the height gradient we also observe a slight increase of the DLSSPW width $w$ that can be attributed to the the proximity effect during EBL.

SPP wavepackets are excited by focusing a TM-polarized laser beam with vacuum wavelength $\lambda=980\,\mathrm{nm}$ onto a grating coupler deposited on top of one (single site excitation, see orange box in Fig.\,\ref{fig:afm_sem} (b)) or several (multi site excitation)  DLSPPWs.
An oil 	immersion objective ($\times 63$ magnification, NA = 1.4) collects the leakage radiation emitted by the SPPs as they propagate along the waveguides.
The real-space SPP intensity distribution $I(x,z)$ is recorded
by imaging the sample plane onto a sCMOS camera (Andor Marana), while 	the corresponding momentum-space intensity distribution $I(k_x,k_z)$ is obtained by imaging the back-focal
plane (BFP) of the oil immersion objective.
According to the quantum optical analogy, theses two quantities are related to the electron probability density $\vert\Psi(x,t)\vert^2$ and the momentum resolved spectrum $\vert\Psi(k,E) \vert^2$, respectively, where $k$ is the quasi-momentum and $E$ is the energy of an electron.

\begin{figure}[t]
	\centering
	\includegraphics[width=\linewidth]{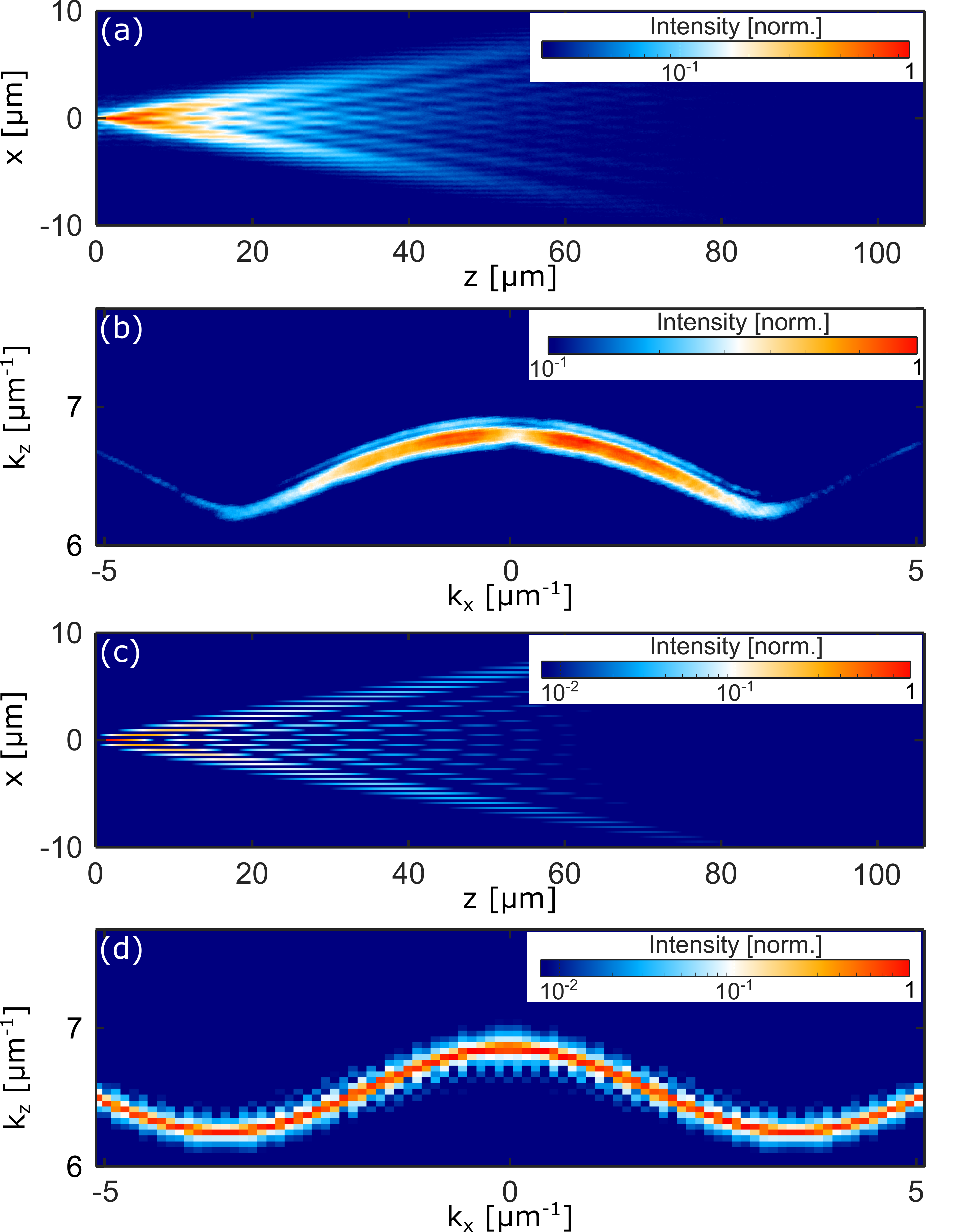}
	\caption{Measured (a) real-space SPP intensity distribution and (b) momentum resolved spectrum of an array of identical DLSPPWs for single site excitation in the center of the array. The corresponding calculated real- and Fourier-space distributions are shown in
		(c) and (d), respectively. }
	\label{fig2}
\end{figure}

We start our discussion with the field-free case, i.e, a periodic lattice with equal on-site energies. 
For this purpose, we have prepared an array of nominally identical DLSPPWs (height $h=140\,\mathrm{nm}$, width $w=300\,\mathrm{nm}$) with constant center-to-center separation $d=900\,\mathrm{nm}$ between neighboring waveguides.
Figure\,\ref{fig2} (a) depicts the real-space leakage radiation intensity distribution for single site excitation in the center of the array.
The  conical spread of the wavepacket as well as the characteristic interference pattern are indicative for discrete diffraction\,\cite{christodoulides2003discretizing}.
This intensity distribution is the optical analogue of the two-state quantum random walk
probability distribution.
The corresponding momentum resolved spectrum recorded by Fourier-space leakage radiation microscopy is shown in Fig.\,\ref{fig2} (b). 
It shows a cosine-like feature that can be interpreted as the tight-binding band of a lattice composed of identical sites with constant couplings. A slight deviation of the band curvature from a perfect cosine shape is due to non-vanishing next-nearest neighbour coupling~\cite{cherpakova2017transverse}.

In order to compare the experiments with theory, we numerically solved\,\cite{Lin2007spatial}
the set of coupled mode equations\,(\ref{CMT:equation}) for an array of $N=61$ equally spaced identical waveguides
and assume $C=0.15\,\mathrm{\upmu m}^{-1}$ and $\beta=(6.5+0.013\imath)\,\mathrm{\upmu m}^{-1}$.
Figures\,\ref{fig2} (c) and (d) show the calculated real- and Fourier-space intensity distributions, respectively, for a single site excitation.
The numerical results qualitatively reproduce the experimentally observed trends. In particular, they show a ballistic spreading of the wavepacket in the real-space data and a cosine-shaped band in the momentum resolved spectrum. 

\begin{figure}[t]
	\centering
	\includegraphics[width=\linewidth]{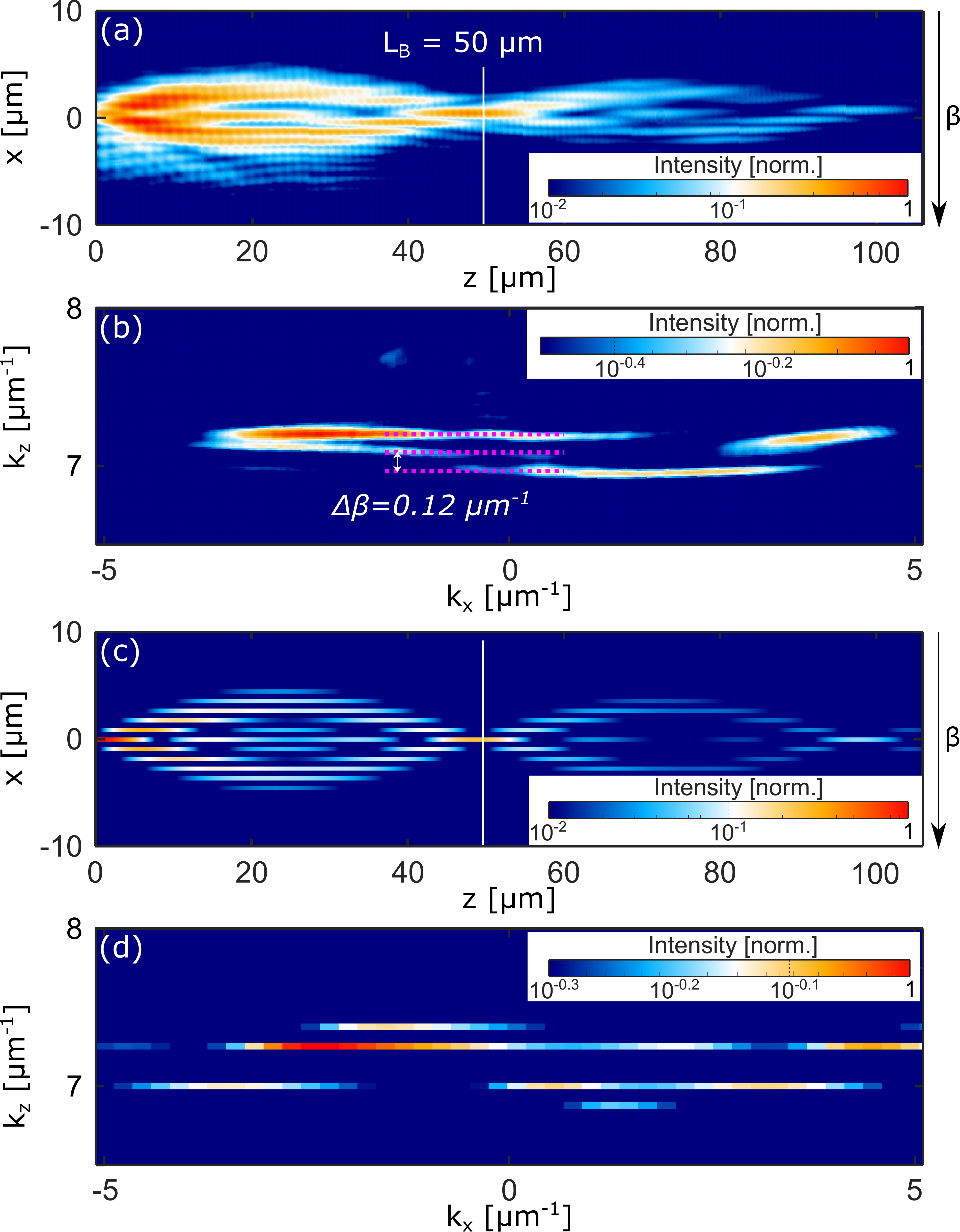}
	\caption{
		Measured (a) real-space SPP intensity distribution and (b) momentum resolved spectrum of an array of 29 DLSPPWs with a gradient of the DLSPPW heigh for single site excitation in the center of the array. The corresponding calculated real- and Fourier-space distributions are shown in (c) and (d), respectively.}
	\label{fig:Fig3}
\end{figure}

Next, we introduce a gradient of the DLSPPW height by linearly varying the dose during the EBL process in order to mimic a DC electric field applied to a tight-binding lattice.
A scanning electron micrograph of the array as well as the height profile of the DLSPPWs are shown in Fig.\,\ref{fig:afm_sem} (b) and (c), respectively.
The real-space SPP intensity distribution for single waveguide excitation in the middle of the array is depicted in Fig.\,\ref{fig:Fig3} (a). 
In contrast to the conical divergence in the field-free case, we observe here a periodic breathing of the wave packet, where the intensity is refocused after a propagation distance $L_B$ of approximately $50\,\mathrm{\upmu m}$. 
The corresponding momentum resolved spectrum (see Fig.\,\ref{fig:Fig3} (b)) features a series of modes with constant $k_z$ separated by an offset $\Delta \beta=0.12\pm0.01\,\mathrm{\upmu m}^{-1}$.
The vanishing group velocity $v_g=dk_z/dk_x=0$ of these modes indicates that they are spatially localized.
These observations are in agreement with the predictions of the coupled mode theory for the propagation of light in an array of evanescently coupled waveguides with linearly varying propagation constants\,\cite{peschel1998optical}.
The eigenmode spectrum of such an array is composed of a series of spatially localized Wannier-Stark states with constant propagation constant offset $\Delta \beta$.
In the case of single-site excitation,  the wavepacket is calculated to periodically refocus after the  Bloch-Oscillation-length \begin{align}
	L_\text{B}=\frac{2\pi}{\Delta\beta}. \label{eq:L_Bloch}
\end{align}
We note for later reference, that the oscillation period of a Gaussian wavepacket in such an array is also given by $L_\text{B}$.
For the measured mode offset $\Delta \beta=0.12\,\mathrm{\upmu m}^{-1}$, the expected Bloch-Oscillation-length is $52\,\mathrm{\upmu m}$, which is in good agreement with the value of $L_B=50\pm2\,\mathrm{\upmu m}$ extracted from the real-space intensity data. 
Based on this analysis, we identify the discrete states observed in the momentum resolved spectrum as steps of the Wannier-Stark ladder.
To further support this interpretation, we numerically solved the coupled mode equations for  a waveguide array with
linearly increasing propagation constants and single site excitation.
The real- and Fourier space intensity distribution presented in Fig.\,\ref{fig:Fig3} (c) and (d), respectively, were calculated using the parameters $C=0.15\,\mathrm{\upmu m}^{-1}$ and 
$\beta_m=\beta_0+m\Delta\beta$, where  $\beta_0=(5.1 + 0.012\imath)\,\mathrm{\upmu m}^{-1}$ and $\Delta\beta=(0.128+0.00015 \imath)\,\mathrm{\upmu m}^{-1}$. 
The comparison of the simulated intensity distributions with the corresponding experimental data shows a good agreement. 

\begin{figure}[t]
	\centering
	\includegraphics[width=\linewidth]{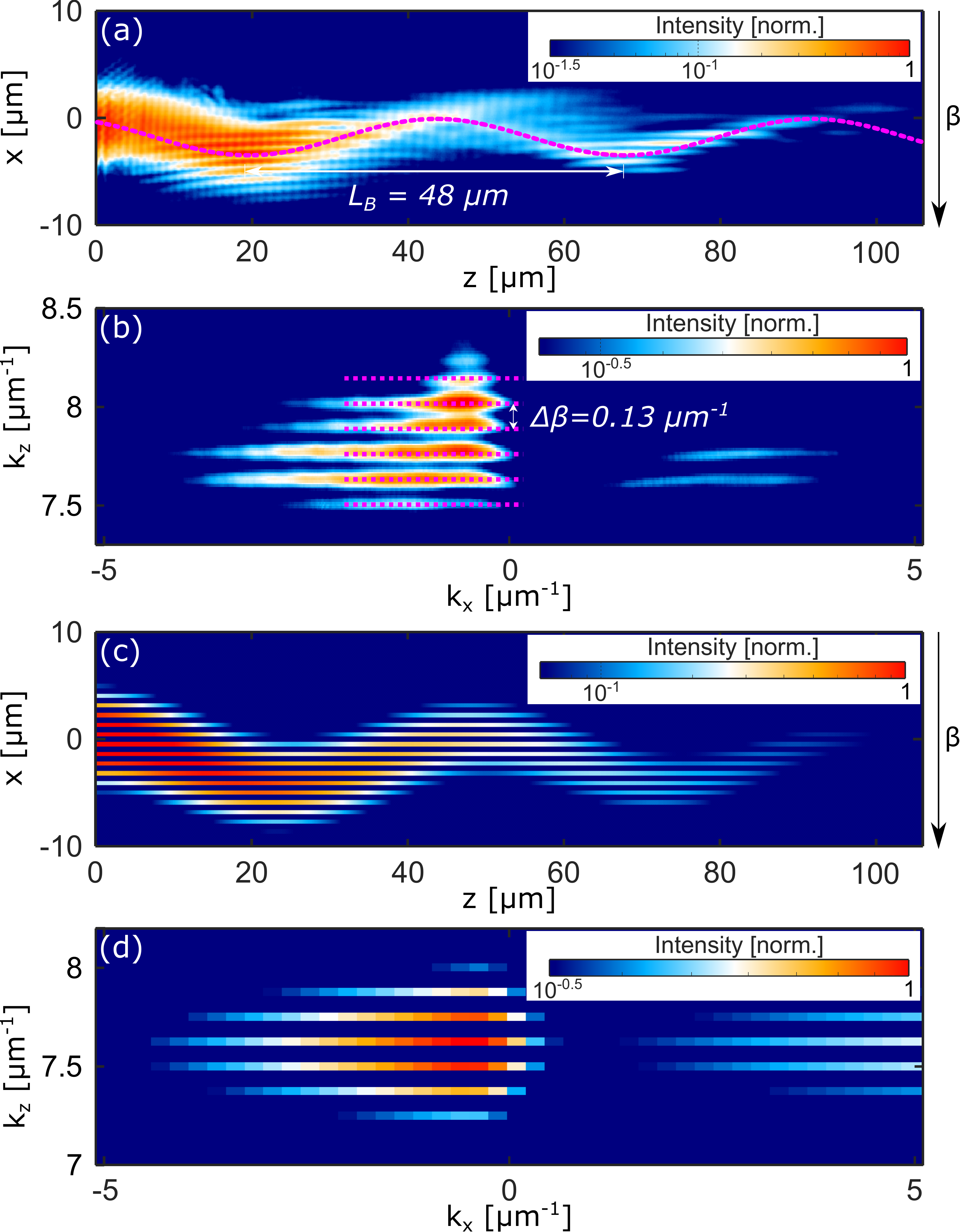}
	\caption{
		Measured (a) real-space SPP intensity distribution and (b) momentum resolved spectrum of an array of 29 DLSPPWs with a gradient of the DLSPPW height for Gaussian excitation of multiple waveguides in the center of the array. The corresponding calculated real- and Fourier-space distributions are shown in (c) and (d), respectively.
	}
	\label{fig:gaussian}
\end{figure} 

To simultaneously excite multiple waveguides with a Gaussian amplitude distribution, we fabricated a waveguide array with  all the same parameters as before except for the grating couplers being on top of all waveguides.
For the chosen laser spot size, we excite a wavepacket with substantial weight in about seven waveguides.
Fig.\,\ref{fig:gaussian} (a) shows the measured real-space intensity  distribution of an array that has nominally the same height gradient as the array discussed in connection with the single site excitation. 
Instead of the periodic breathing of the intensity distribution observed in the previous case, the excitation of multiple waveguides results in a periodic oscillatory motion with an oscillation period $L_B=48\pm1\,\mathrm{\upmu m}$.
The oscillatory motion of the wave packet in real-space is the plasmonic analogue of the Bloch oscillation of an electron in the related lattice system.
Fig.\,\ref{fig:gaussian} (b) depicts the corresponding momentum resolved spectrum. 
It is composed of a set of seven discrete modes with constant $k_z$ separated by a constant offset $\Delta \beta=0.13\pm0.005\,\mathrm{\upmu m}^{-1}$.
In the light of the previous discussion and in accordance with theory\,\cite{peschel1998optical}, we can identify the discrete modes in the spectrum as localized Wannier-Stark states.
In comparison to single site excitation, a larger number of Wannier-Stark states is observed in case of multiple waveguide excitation (compare Fig.\,\ref{fig:Fig3}(b) and Fig.\,\ref{fig:gaussian}(b)).
This observation confirms the localized character of the Wannier-Stark states.

Figures \ref{fig:gaussian}(c) and (d) depict the corresponding CMT simulated real-space intensity distribution and the momentum resolved spectrum, respectively, for the previous given parameters $C=0.15\,\mathrm{\upmu m}^{-1}$ and 
$\beta_m=\beta_0+m\Delta\beta$, where  $\beta_0=(5.1 + 0.012\imath)\,\mathrm{\upmu m}^{-1}$ and $\Delta\beta=(0.128+0.00015 \imath)\,\mathrm{\upmu m}^{-1}$. 
We again find that the calculations qualitatively reproduce all experimentally
observed trends.

We note that the excited Wannier-Stark states have the strongest spectral weight at negative $k_x$ (see Fig.\, \ref{fig:gaussian}(b) and (d)). This effect can be explained by the propagation losses in the array. The excited wavepacket first moves in direction of increasing $\beta$ (negative $x$ direction). After the distance $L_B/2$, it reaches the turning point and then moves in positive $x$-direction.  Due to losses, the intensity in the second half of the oscillation is lower than in the first half. This explains the asymmetry of the momentum resolved spectrum with respect to $k_x$.

In conclusion, we investigated the propagation of SPPs in arrays of coupled DLSPPWs.
In the case of an array with identical waveguides and single site excitation, the  wavepacket exhibits in real-space  ballistic spreading while the corresponding spectrum features a continuous band. 
The introduction of a gradient of the propagation constant alters the wave dynamics.
For single (multiple) site exciation, we observe a periodic breathing (oscillation) of the wavepacket.
The corresponding spectra feature a set of equally spaced discrete modes that we identify as the steps of the Wannier-Stark ladder.
The experimental results are in good agreement with numerical calculations based on the couple mode theory.

We acknowledge financial support by the Deutsche Forschungsgemeinschaft through 
CRC/TR 185 (277625399) OSCAR.
\bibliography{sample}

\end{document}